\documentclass[12pt]{article}
\newcommand {\evis} {$E_{\mathrm{vis}}$}
\newcommand{\dmsq}{$\Delta m^2$}

\newcommand {\omue} {$\nu_\mu \rightarrow \nu_e$}
\newcommand {\omuebar} {$\bar{\nu}_\mu \rightarrow \bar{\nu}_e$}
\def\nuequasi{$\nu_e n \rightarrow e^- p$}
\def\anuequasi{$\bar{\nu}_e p \rightarrow e^+ n$}
\def\nue_pizero{$\nu_e n \rightarrow e^- \pi^0 p$}
\def\anue_pizero{$\bar{\nu}_e p \rightarrow e^+ \pi^0 n$}
\def\ypi_gg{$\pi^0 \rightarrow \gamma\gamma$}

\begin {document}
\title
{A non-QE signature of $\nu_e$ appearance in a water Cherenkov detector}
\author
{A. Asratyan and V. Verebryusov\\
ITEP Moscow, Russia}
\date {\today}
\maketitle

\begin{abstract}
We argue that analyzing the $\nu_e$-induced CC reaction \nue_pizero\ 
along with the quasielastic reaction \nuequasi\  may significantly 
enhance the sensitivity of a water Cherenkov detector to the subleading 
oscillation \omue\ at neutrino energies $\sim$ 2--3 GeV, as projected for 
the off-axis neutrino beam of NuMI. At the level of standard selections, 
the multi-ring signal of \omue\ yielded by this $\pi^0$-producing reaction 
is comparable to the 1-ring (quasielastic) signal in statistical 
significance. The neutral-current background to \nue_pizero\ can be 
further suppressed by analyzing spatial separation between the 
reconstructed primary vertex and the vertices of individual rings.
\end{abstract}

     Detecting the "subleading" oscillation \omue\ in an off-axis beam 
with peak energy near 2--3 GeV has emerged as one of the major goals
of the NuMI program \cite{numi}, addressed at a Fermilab workshop in
May 2002 \cite{workshop} and in a recent Letter of Intent on the subject 
\cite{loi}. A 2.5--3 times bigger baseline than in the proposed JHF2K 
experiment \cite{jhf2k}, that will operate at a lower beam energy of 
$E_\nu \sim$ 0.7--0.8 GeV, will allow to probe the matter effect by 
comparing the probabilities of the \omue\ and \omuebar\ 
transitions\footnote{Without antineutrino 
running, this can be done by comparing
the \omue\ probabilities for a longer 
(NuMI) and shorter (JHF2K) baselines.}.
Detector options discussed include a fine-grained low-density 
calorimeter, a liquid-Argon TPC, and a water Cherenkov spectrometer. 
Of these, the latter option is based on proven techniques,
has an excellent record in neutrino physics \cite{atmo}, and offers best
opportunities in terms of maximum target mass at reasonable cost. It 
appears however that, in the quasielastic mode only discussed thus far, 
a water Cherenkov detector will perform worse in NuMI than in 
JHF2K because of more background to 1-ring electronlike events from 
single-$\pi^0$ production in NC collisions \cite{loi}. It has been
estimated \cite{deborah} that, due to considerable NC background to
quasielastics, a water Cherenkov detector needs to be $\sim 3$ times as 
massive as a low-density calorimeter in order to reach similar     
sensitivity to \omue\ in NuMI conditions. But can the performance of a 
water Cherenkov detector at $E_\nu \sim$ 2--3 GeV be boosted by going 
beyond quasielastics ?

\begin{figure}
\vspace{18 cm}
\includegraphics{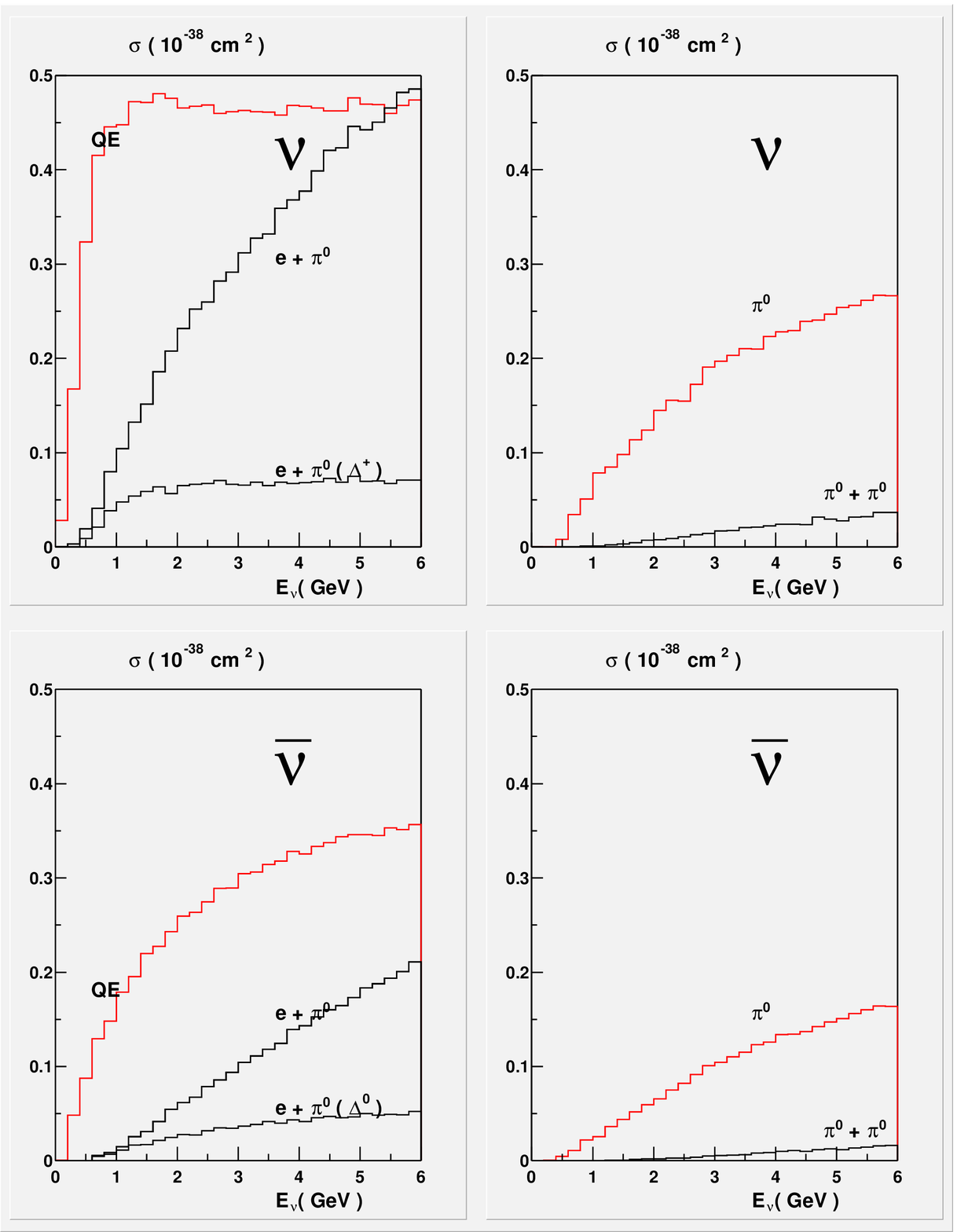}
\caption
{Cross sections per average nucleon in water of the reactions 
\nuequasi\ and \nue_pizero\ (top left), \anuequasi\ and \anue_pizero\
(bottom left),
$\nu N \rightarrow \nu \pi^0 N$ and
$\nu N \rightarrow \nu \pi^0 \pi^0 N$ (top right), and
$\bar{\nu} N \rightarrow \bar{\nu} \pi^0 N$ and
$\bar{\nu} N \rightarrow \bar{\nu} \pi^0 \pi^0 N$ (bottom right)
as functions of neutrino energy. Also shown are the contributions of
$\Delta(1232)$ excitation to the \nue_pizero\ and \anue_pizero\ cross
sections.}
\label{xsections}
\end{figure}

     We believe that at neutrino energies $\sim 2$ GeV or higher,
the sensitivity to \omue\ can be enhanced by also detecting the CC 
reactions producing a $\pi^0$, \nue_pizero\ and \anue_pizero, that 
largely proceed through excitation
of the $\Delta(1232)$ and other baryon 
resonances\footnote{The first observation in
a water Cherenkov detector of the 
corresponding $\nu_\mu$-induced reaction,
$\nu_\mu n \rightarrow \mu^- \pi^0 p$,
has been reported in \cite{first}.}.
Compared to \nuequasi\ and \anuequasi, the cross sections of these 
reactions\footnote{The 
cross sections of CC and NC 
reactions are quoted according 
to NEUGEN \cite{neugen}.}
are small for $E_\nu < 1$ GeV but significant at $E_\nu \sim$ 2--3 GeV 
(see Fig. \ref{xsections}), so that these processes may be relevant to 
NuMI rather than JHF2K. Depending on whether or not the $\pi^0$ 
has been fully reconstructed, two different signatures are possible for 
a water Cherenkov detector:
\begin{itemize}
\item
Three $e$-like rings, of which two fit to \ypi_gg;
\item
Two $e$-like rings that would not fit to a $\pi^0$.
\end{itemize}
Despite a smaller cross section, the $\pi^0$-producing reaction may be 
competitive with quasielastics because of less neutral-current background:
two $\pi^0$ mesons, and not just one, have to be produced in order to 
mimic the aforementioned 2- and 3-ring signatures. At neutrino energies 
of a few GeV in particular, the cross section of the NC reaction
$\nu N \rightarrow \nu \pi^0 \pi^0 N$
should be kinematically suppressed compared to
$\nu N \rightarrow \nu \pi^0 N$. 
This simple conjecture is supported by NEUGEN predictions\footnote{The 
uncertainties of these predictions 
for the cross sections of \nue_pizero\ 
and $\nu N \rightarrow \nu \pi^0 \pi^0 N$ 
are hard to estimate, as the data 
are scarce for the former reaction and 
totally lacking---for the latter
reaction.},
see Fig. \ref{xsections}.

\begin{figure}
\vspace{18 cm}
\includegraphics{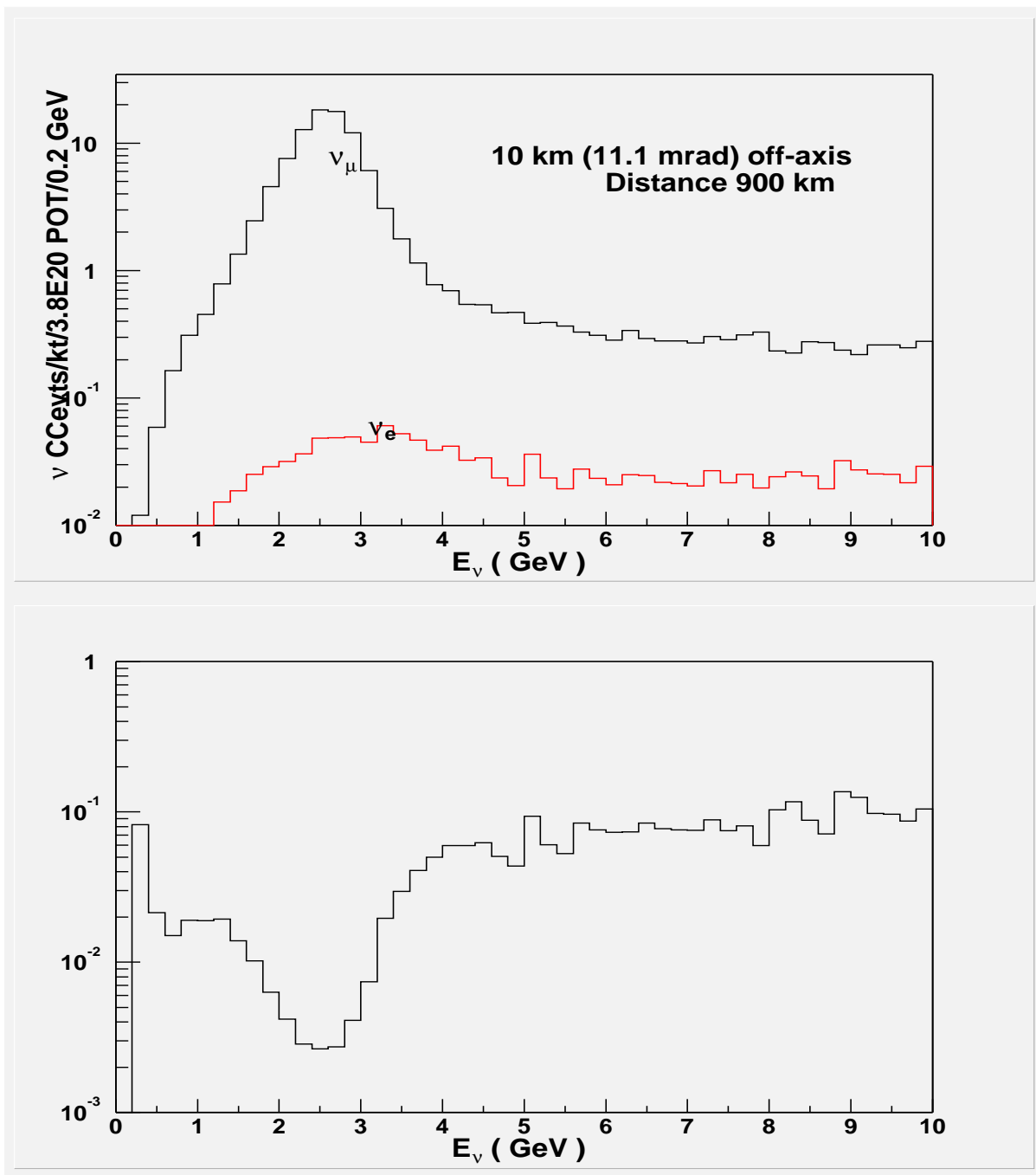}
\caption
{The oscillation-free $E_\nu$ spectra of $\nu_e$- and $\nu_\mu$-induced 
CC events (top panel) and their ratios (bottom panel) for an off-axis 
location in the NuMI medium-energy beam ($L = 900$ km and $R = 10$ km). 
The exposure is 100 kton--years.}
\label{all_cc}
\end{figure}

     The values of oscillation parameters  assumed in the simulation are 
\dmsq\ = 0.003 eV$^2$, $\sin^2 2\Theta_{23} = 1$, and 
$\sin^2 2\Theta_{13} = 0.1$ (that is, at the CHOOZ limit \cite{chooz}). 
Matter effects are not accounted for, as here we only wish to compare the 
oscillation signals in the single-ring and multi-ring channels. Apart
from enhancing the matter effect, increasing the baseline shifts the 
oscillation maximum to higher values of $E_\nu$ where the cross sections 
of the reactions \nue_pizero\ and \anue_pizero\ are relatively big. 
Therefore, we select a baseline close to the maximum value for NuMI: 
$L = 900$ km.
The displacement from the axis of the NuMI medium-energy beam \cite{numi},
$R$, is varied in the simulation. The $E_\nu$ distribution 
of all $\nu_\mu$-induced CC events in the absence of oscillations, 
illustrated in Fig. \ref{all_cc} for the neutrino mode and $R = 10$ km,
peaks at $E_\nu \simeq 2.6$ GeV. In the peak region, the intrinsic 
$\nu_e$ component of the beam is some 0.3\% of the $\nu_\mu$ component 
(see Fig. \ref{all_cc}). Running in the antineutrino mode will yield 
some 3 times less CC events for the same number of delivered protons.

     Taking into account the experimental conditions of a water 
Cherenkov detector, actually simulated are the "quasi-inclusive" CC 
reactions
$\nu_e N \rightarrow e^- X$  and
$\nu_e N \rightarrow e^- \pi^0 X$ \footnote{Respective 
antineutrino reactions are implicitly included.}
and flavor-blind NC reactions
$\nu N \rightarrow \nu \pi^0 X$  and
$\nu N \rightarrow \nu \pi^0 \pi^0 X$
in neutrino collisions with water. Here, $X$ denotes a system of hadrons 
other than the $\pi^0$, in which the momenta of all charged particles are 
below the Cherenkov threshold in water. These reactions are analyzed in
terms of visible energy \evis, defined as a sum over the energies of all 
detectable particles: the $\pi^0$ mesons(s) and the charged lepton 
for CC reactions.

\begin{figure}
\vspace{18 cm}
\includegraphics{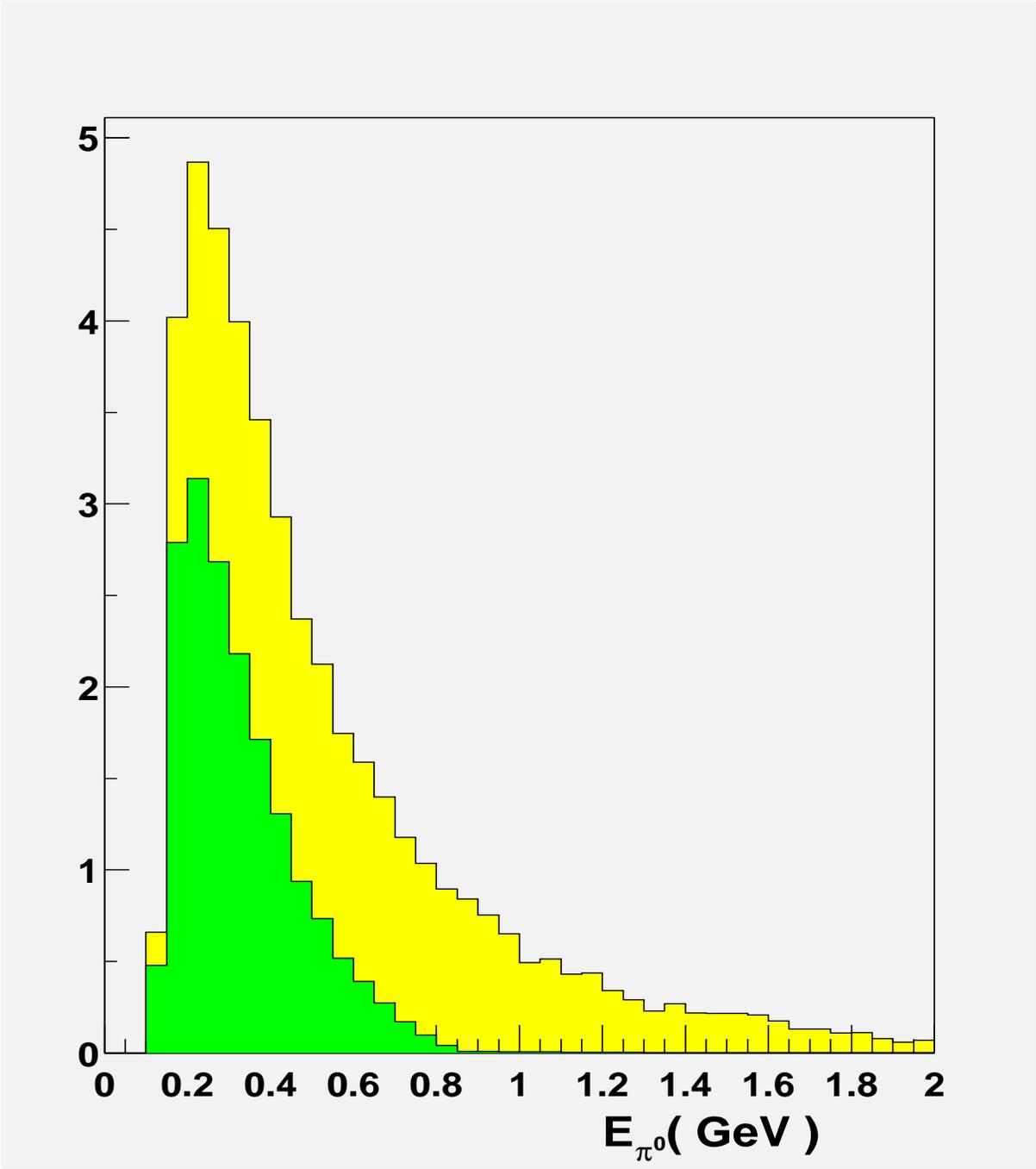}
\caption
{The momenta of $\pi^0$ mesons emitted in the CC reaction
$\nu_e N \rightarrow e^- \pi^0 X$. The lower histogram shows
the contribution of reconstructed $\pi^0$ mesons.}
\label{pimom}
\end{figure}

     In a water Cherenkov detector, the two photons from \ypi_gg\ may
show up as a single $e$-like ring because of a small opening angle (this
largely occurs at high $\pi^0$ momenta), or because one of the photons 
from an "asymmetric" $\pi^0$ decay is too soft to be detected \cite{jhf2k}. 
The efficiency of $\pi^0$ reconstruction as a function of its momentum will 
depend on the geometry and instrumentation of a Cherenkov detector; the 
estimates quoted below are based on the results for the 1-kiloton detector 
of K2K \cite{k2k}, as reported in \cite{pizero}. The momenta of $\pi^0$ 
mesons emitted in $\nu_e N \rightarrow e^- \pi^0 X$ 
are plotted in Fig. \ref{pimom}, that also shows the distribution of 
reconstructed $\pi^0$ mesons (lower histogram). We assume that at least 
one photon from \ypi_gg\ is always detected, so that  all 1-ring CC events 
arise from 
$\nu_e N \rightarrow e^- X$  
and all 1-ring NC events---from 
$\nu N \rightarrow \nu \pi^0 X$ 
with unresolved photon showers. The probability for two photons to form a 
fake $\pi^0$ candidate is neglected (in SuperK, the r.m.s. width of the 
$\pi^0$ peak is only $\sim 40$ MeV \cite{pizero}). Depending on whether or 
not the $\pi^0$ is reconstructible, a CC collision
$\nu_e N \rightarrow e^- \pi^0 X$ 
will produce 3 or 2 rings in the detector. NC events showing 3 (2) rings 
arise from failing to reconstruct one $\pi^0$ (both $\pi^0$s) in the 
reaction 
$\nu N \rightarrow \nu \pi^0 \pi^0 X$.

\begin{figure}
\vspace{18 cm}
\includegraphics{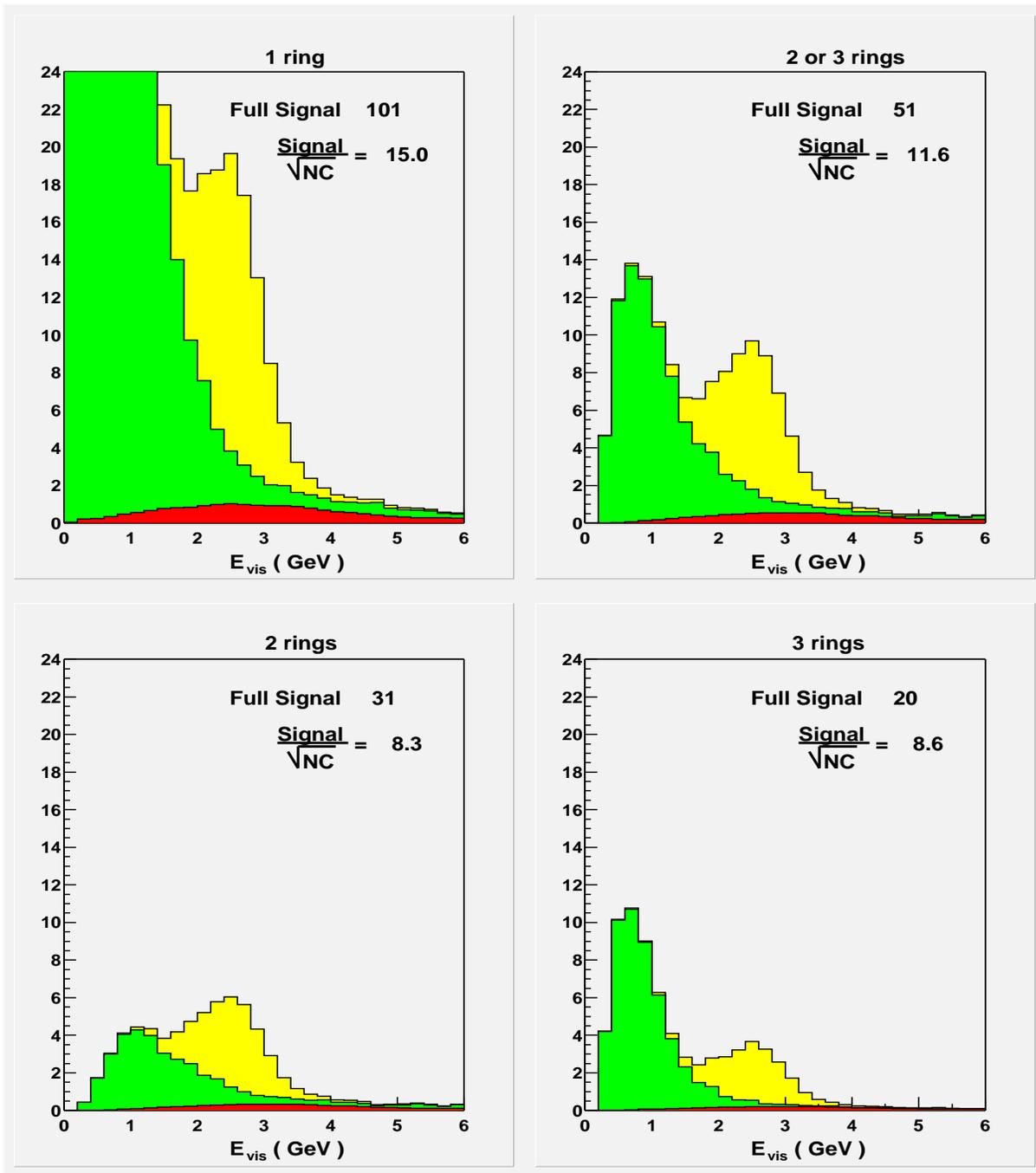}
\caption
{\evis\ distributions of events featuring one $e$-like ring (top left), 
2 or 3 rings (top right), 2 rings (bottom left), and 3 rings (bottom 
right). Here and in subsequent Figures, shown for either event category 
are the \omue\ signal (yellow area), the NC background (green area), and 
the intrinsic CC background (red area). For incident neutrinos
and $R = 9$ km.}
\label{nu_9}
\end{figure}

\begin{figure}
\vspace{18 cm}
\includegraphics{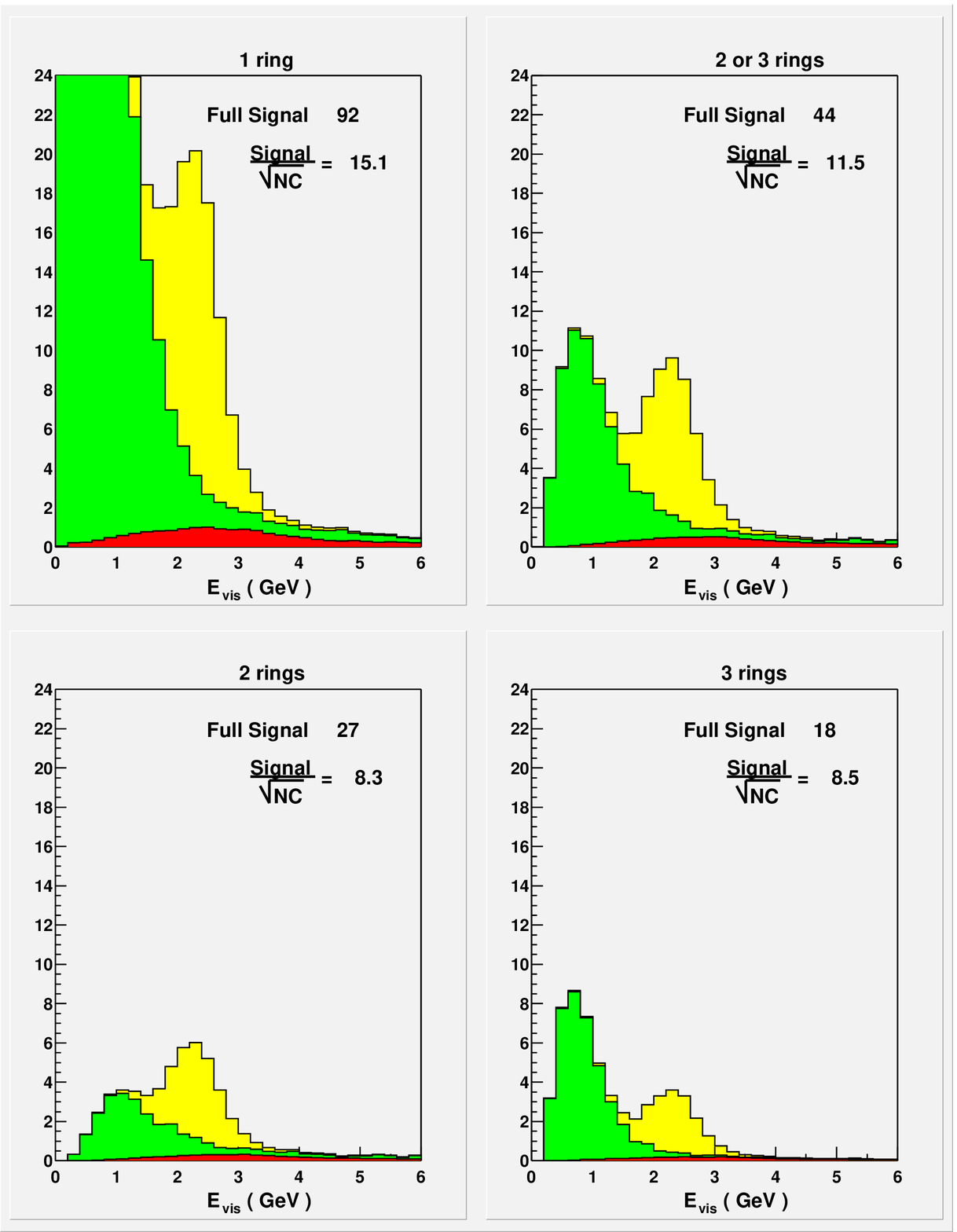}
\caption
{\evis\ distributions of 1-ring and multi-ring events for incident 
neutrinos and $R = 10$ km.}
\label{nu_10}
\end{figure}

\begin{figure}
\vspace{18 cm}
\includegraphics{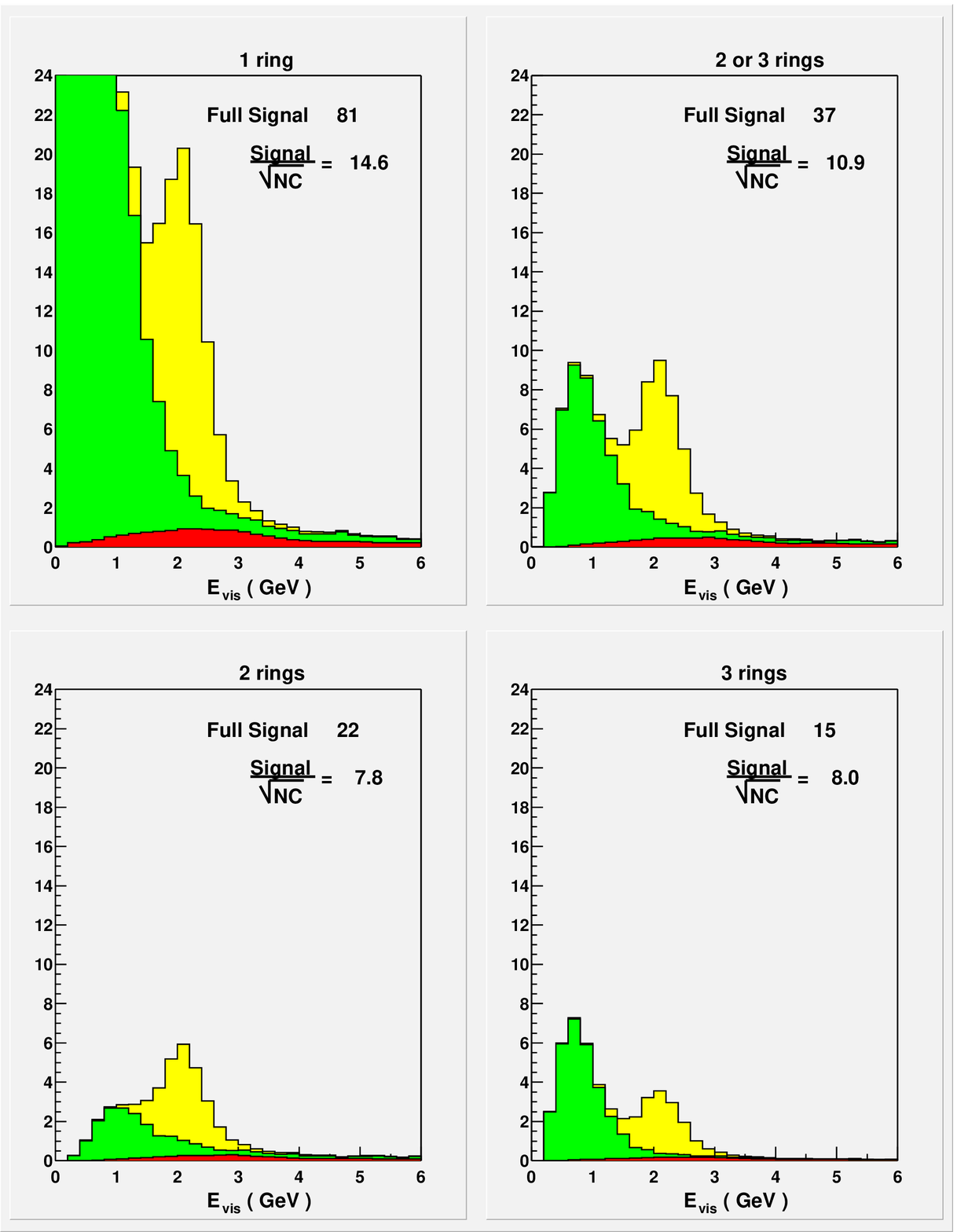}
\caption
{\evis\ distributions of 1-ring and multi-ring events for incident
neutrinos and $R = 11$ km.}
\label{nu_11}
\end{figure}

     The \evis\ distributions\footnote{Failing to
reconstruct a $\pi^0$ will but weakly affect the 
value of visible energy: in this case, either 
the two photons from \ypi_gg\ have merged into a 
single shower sampled as a whole, or one of them 
is very soft.}
of events featuring 1, 2, and 3 $e$-like rings are shown in 
Figs. \ref{nu_9}--\ref{nu_11} for incident neutrinos and different values 
of $R$. The three components of the \evis\ distribution for either channel
are: the \omue\ signal (yellow area), the NC background (green area), and
the intrinsic-$\nu_e$ background (red area). The \evis\ interval for 
estimating the effect is selected so as to maximize the "Figure of Merit" 
$S/\sqrt{B}$, where $S$ is the number of signal events and $B$ is the 
total (NC plus intrinsic-CC) background. For either the $\nu$ and 
$\bar{\nu}$ settings of the beam, Table \ref{stati_1} compares the 1-ring 
and multi-ring samples in terms of total \omue\ signals, numbers of signal
and background events in the selected \evis\ windows, and statistical
significance. Predictably, the ratio between the multi-ring and 1-ring 
signals decreases with increasing $R$ (or off-axis angle). For incident 
neutrinos, the multi-ring signal is $\sim$2 times less than the 1-ring 
signal in absolute value, but has comparable significance due to less NC 
background. For incident antineutrinos, the multi-ring signal is 
substantially less significant than the 1-ring signal.

\begin{table}[h]
\begin{tabular}{|c|c|c|c|c|c|c|}
\hline
\hline
Beam, radius,& Total & \evis\ & Signal in & NC    & Intr. CC&$S/\sqrt{B}$\\
signature    & signal& window & window    &backgr.& backgr. & (FoM)   \\
\hline
\hline
$\nu$, $R = 9$ km:  & & & & & & \\
1 ring       & 101.&2.2--3.2 GeV& 61.    & 11.6     & 4.8    &  15.0  \\
2 or 3 rings & 51. &2.0--3.4 GeV& 39.    &  7.5     & 3.6    &  11.6  \\
\hline
$\nu$, $R = 10$ km:  & & & & & & \\
1 ring       &  92.&2.0--3.0 GeV& 60.    & 11.0     & 4.7    &  15.1  \\
2 or 3 rings &  44.&2.0--3.0 GeV& 30.    &  4.2     & 2.4    &  11.5  \\
\hline
$\nu$, $R = 11$ km:  & & & & & & \\
1 ring       &  81.&1.8--2.8 GeV& 57.    & 10.5     & 4.4    &  14.6  \\
2 or 3 rings &  37.&1.6--2.8 GeV& 31.    &  5.6     & 2.5    &  10.9  \\
\hline
\hline
$\bar{\nu}$, $R = 9$ km:  & & & & & & \\
1 ring       &  80.&2.0--3.2 GeV& 56.    &  5.0     & 4.9    &  17.9  \\
2 or 3 rings &  25.&2.0--3.2 GeV& 17.    &  2.3     & 1.7    &   8.5  \\
\hline
$\bar{\nu}$, $R = 10$ km:  & & & & & & \\
1 ring       &  69.&1.8--3.0 GeV& 53.    & 5.1      & 4.5    &  17.0  \\
2 or 3 rings &  20.&1.8--2.8 GeV& 14.    & 1.9      & 1.2    &  7.8  \\
\hline
$\bar{\nu}$, $R = 11$ km:  & & & & & & \\
1 ring       &  59.&1.8--2.6 GeV& 38.    & 3.2     & 2.8    &  15.3  \\
2 or 3 rings &  16.&1.6--2.8 GeV& 12.    & 2.1     & 1.3    &   6.7  \\
\hline         
\hline
\end{tabular}

\caption
{The total \omue\ (\omuebar) signal and the numbers of signal, NC 
background, and intrinsic-CC background events in the selected \evis\ 
window for 1-ring and multi-ring signatures and for the $\nu$ and 
$\bar{\nu}$ settings of the beam. Also quoted is the "Figure of Merit"
$S / \sqrt{B}$, where $S$ is the number of signal events and $B$ is
the total (NC plus intrinsic-CC) background. The assumed exposure is 100 
kton--years.}
\label{stati_1}
\end{table}

     In a realistic Cherenkov detector, recoil protons often escape
detection even for momenta above the Cherenkov threshold \cite{svoboda}.
On average, recoil protons have higher momenta in \nue_pizero\ than in
\nuequasi\ due to a broader $Q^2$ distribution, so that the multi-ring
signal is expected to benefit most from keeping (some) energetic protons.
That lifting the upper cut on proton momentum effectively increases the
ratio between the multi-ring and 1-ring signals is illustrated by 
Table \ref{stati_2}, to be compared with Table \ref{stati_1}.

\begin{table}[h]
\begin{tabular}{|c|c|c|c|c|c|c|}
\hline
\hline
Beam, radius,& Total & \evis\ & Signal in & NC    & Intr. CC&$S/\sqrt{B}$\\
signature    & signal& window & window    &backgr.& backgr. & (FoM)   \\
\hline
\hline
$\nu$, $R = 9$ km:  & & & & & & \\
1 ring       & 133.&2.0--3.4 GeV& 87.    & 24.9     & 8.7    &  14.9  \\
2 or 3 rings &  80.&2.0--3.2 GeV& 49.    & 9.9     & 5.4    &  12.4  \\
\hline
$\nu$, $R = 10$ km:  & & & & & & \\
1 ring       & 119.&1.8--3.0 GeV& 80.    & 22.0     & 7.2    &  14.9  \\
2 or 3 rings &  68.&1.6--3.0 GeV& 51.    & 12.3     & 5.6    &  12.1  \\
\hline
$\nu$, $R = 11$ km:  & & & & & & \\
1 ring       & 105.&1.8--2.6 GeV& 58.    & 12.3     & 4.6    &  14.1  \\
2 or 3 rings &  56.&1.6--2.6 GeV& 37.    & 7.2     & 3.6    &  11.3  \\
\hline
\hline
\end{tabular}

\caption
{The 1-ring and multi-ring signals of \omue\ compared for incident
neutrinos, no longer requiring that proton momenta be below the 
Cherenkov threshold in water.}
\label{stati_2}
\end{table}

     As indicated in \cite{loi}, fast PMT's and good
photocathode coverage may help discriminate between the electron- and
$\pi^0$-induced showers by detecting the spatial separation between the
conversion points of the two photons from \ypi_gg. If shown to be
realistic, this will equally apply to 1-ring and multi-ring signatures
of \omue. Yet another geometric handle may be possible for multi-ring 
topologies only, provided that spatial
resolution of the detector is better than photon conversion length
$\lambda_c$. An important advantage of having more than one ring 
is that constraining the axes of all rings to a common point in space will 
yield the position of the primary vertex. Within errors, this should 
coincide with the reconstructed vertex of a $e^-$-induced shower, whereas 
the vertex of an unresolved $\pi^0$ shower will be displaced by $\sim 
\lambda_c$ along
the shower direction. The spatial resolution of SuperK has been estimated 
as 18 cm for the vertex of proton decay $p \rightarrow e^+ \pi^0$ whose 
signature is very similar to that of $\nu_e n \rightarrow e^- \pi^0 p$, 
and as 34 cm for the vertex of a single $e$-like ring \cite{resolution}.
We have $\lambda_c \simeq 40$ cm for water, so that even a modest 
improvement in resolution over SuperK will allow to efficiently 
discriminate between CC and NC multi-ring events and to measure the NC 
background (this of course needs to be checked by a detailed simulation 
of detector response).

     To conclude, our preliminary results indicate that analyzing the
reaction \nue_pizero\ along with the quasielastic reaction \nuequasi\ 
may significantly enhance the sensitivity of a water Cherenkov detector 
to the "subleading" oscillation \omue\ at neutrino energies 
$\sim$ 2--3 GeV. At the level of standard selections, the statistical 
significance of the multi-ring signal of \omue\ is comparable to that 
of the 1-ring (quasielastic) signal. The antineutrino reaction 
\anue_pizero\ is a less efficient probe of \omuebar\ because its cross 
section is small. The NC background to \nue_pizero\ can be suppressed by 
reconstructing the vertex of neutrino collision and analyzing spatial 
separation between the primary and secondary vertices.

\clearpage

\end{document}